# Ubiquitous Spin Freezing in the Superconducting State of UTe$_2$

Shyam Sundar[1,3], Nasrin Azari[1], Mariah R. Goeks[1], Shayan Gheidi[1], Mae Abedi[1], Michael Yakovlev[1], Sarah R. Dunsiger[1,2], John M. Wilkinson[4], Stephen J. Blundell[4], Tristin E. Metz[5], Ian M. Hayes[5], Shanta R. Saha[5], Sangyun Lee[6], Andrew J. Woods[6], Roman Movshovich[6], Sean M. Thomas[6], Nicholas P. Butch[5,7], Priscila F. S. Rosa[6], Johnpierre Paglione[5,8], Jeff E. Sonier[1,*]

[1]Department of Physics, Simon Fraser University, Burnaby, British Columbia V5A 1S6, Canada

[2]Centre for Molecular and Materials Science, TRIUMF, Vancouver, British Columbia V6T 2A3, Canada

[3]Instituto de Fisica, Universidade Federal do Rio de Janeiro, 21941-972 Rio de Janeiro, RJ, Brazil

[4]Clarendon Laboratory, University of Oxford, Department of Physics, Oxford OX1 3PU, United Kingdom

[5]Maryland Quantum Materials Center, Department of Physics, University of Maryland, College Park, Maryland 20742, USA

[6]Los Alamos National Laboratory, Los Alamos, New Mexico 87545, USA

[7]NIST Center for Neutron Research, National Institute of Standards and Technology, Gaithersburg, Maryland 20899, USA

[8]Canadian Institute for Advanced Research, Toronto, Ontario M5G 1Z8, Canada

[*]email: jsonier@sfu.ca

**Abstract**

In most superconductors electrons form Cooper pairs in a spin-singlet state mediated by either phonons or by long-range interactions such as spin fluctuations. The superconductor $UTe_2$ is a rare material wherein electrons are believed to form pairs in a unique spin-triplet state with potential topological properties. While spin-triplet pairing may be mediated by ferromagnetic or antiferromagnetic fluctuations, experimentally, the magnetic properties of $UTe_2$ are unclear. By way of muon spin rotation/relaxation (μSR) measurements on independently grown $UTe_2$ single crystals we demonstrate the existence of magnetic clusters that gradually freeze into a disordered spin frozen state at low temperatures. Our findings suggest that inhomogeneous freezing of magnetic clusters is linked to the ubiquitous residual linear term in the temperature dependence of the specific heat ($C$) and the low-temperature upturn in $C/T$ versus $T$. The omnipresent magnetic inhomogeneity has potential implications for experiments aimed at establishing the intrinsic low-temperature properties of $UTe_2$.

**Introduction**

The Cooper pairs that form at the onset of superconductivity can be in a state of total spin $S = 0$ (spin singlet) or $S = 1$ (spin triplet). Although extremely rare in a solid-state material, spin-triplet pairing is of both fundamental and practical interest. With spin-aligned Cooper pairs, spin-triplet superconductors naturally possess the crucial element required for realizing superconducting spintronics characterized by a non-dissipative spin current[1,2]. Moreover, spin-triplet superconductors of odd parity, in which the spatial component of the superconducting wavefunction is an odd function, are intrinsically topological and as such support Majorana fermions[3]. Encoding of topologically protected quantum information in Majorana-based qubits is considered a promising and perhaps the only way of achieving a quantum computer robust to environmental noise[4].

The discovery of low-temperature superconductivity in $UTe_2$[5] followed nearly three decades of research on $Sr_2RuO_4$, another low superconducting transition temperature ($T_c$) compound[6] and the first strong candidate for a topological spin-triplet superconductor[7,8]. Yet, recent precision experiments have challenged the longstanding belief that $Sr_2RuO_4$ is an odd-parity spin triplet superconductor[9,10,11], thus heightening the interest in $UTe_2$. There are several experimental indications of spin-triplet Cooper pairing in UTe$_2$, including a negligible (small) decrease in the temperature dependence of the local spin susceptibility in the superconducting state, as inferred from $^{125}$Te-nuclear magnetic resonance (NMR) Knight shift measurements for a magnetic field applied along the $a$ ($b$ or $c$) axis[12,13]. Despite having a low $T_c$ value near 2 K, UTe$_2$ also exhibits a large anisotropic upper critical field $H_{c2}$ that greatly exceeds the upper limit in conventional spin-singlet superconductors from Cooper pair breaking due to the Pauli paramagnetism of the electrons[5,14]. Furthermore, well above $H_{c2}$ superconductivity re-emerges in UTe$_2$ for magnetic fields as high as 60 T applied in specific crystallographic directions[15,16].

The observation of chiral in-gap surface states by scanning tunnelling spectroscopy suggests that UTe$_2$ is a spin-triplet superconductor with a nontrivial band topology[17]. Based on observations of a double phase transition in the specific heat at ambient pressure and a finite polar Kerr effect below $T_c$, which is indicative of a net magnetization perpendicular to the surface, UTe$_2$ is proposed to have a two-component superconducting order parameter that breaks time reversal symmetry[18]. Weyl nodes, where valence and conduction bands cross at a single point near the Fermi surface can occur in the superconducting gap for the possible two-component order parameters inferred by these data, bolstering the case for topological superconductivity in UTe$_2$.

Despite these developments many questions remain, including a critical one related to the underlying pairing mechanism responsible for superconductivity. Initially UTe$_2$ was believed to be the end member of the series of ferromagnetic (FM) uranium-based superconductors UGe$_2$, URhGe and UCoGe, whose spin-triplet Cooper pairing is mediated by FM fluctuations[5]. Unlike the other members in this group wherein the Cooper pair spins align with the internal field generated by pre-existing FM order[19], UTe$_2$ does not exhibit long-range FM order. One scenario is that superconductivity in UTe$_2$ is driven by strong magnetic fluctuations near a FM quantum critical point. This picture is supported by scaling of magnetization data[5] and scaling of the bulk magnetic susceptibility with the dynamic spin susceptibility[20] above $T_c$ for a field applied along the easy axis of magnetization (the crystallographic $a$-axis) that agree with predictions for a three-dimensional (3-D) itinerant-electron system close to a FM instability. Analogous scaling of zero-field μSR data[21] down to ~ 0.4 K is also suggestive of low-frequency FM spin fluctuations that persist and coexist with superconductivity, and magnetic-field-trained polar Kerr effect measurements detect the presence of a FM susceptibility[22]. In contrast, inelastic neutron scattering (INS) experiments on UTe$_2$ were first interpreted as a sign of dominant low-dimensional antiferromagnetic (AFM) spin fluctuations, along with an absence of FM fluctuations[23,24]. Subsequent work has shown that the energy, momentum, and temperature dependence of these low-energy AFM-like magnetic excitations arises from the heavy electron band structure, and that it is not possible based on these excitations alone to conclude whether FM or AFM interactions are dominant[25]. In the superconducting state, a magnetic excitation near 1 meV, at finite momentum transfer, is accompanied by the apparent opening of a spin gap, which can be interpreted either as an exotic superconducting resonance or a response of the underlying magnetic fluctuations to the emergence of superconductivity[25,26,27]. Theoretically, FM fluctuations along the *a*-axis are predicted at ambient pressure for itinerant *f* electrons in UTe$_2$[28]. By contrast, for localized *f* electrons the two-leg ladder-type arrangement of the U atoms may suppress long-range magnetic order by geometrical frustration and coexisting FM and AFM fluctuations at ambient pressure and zero field are possible[29]. While spin-triplet pairing may arise from either FM or AFM[30] spin fluctuations, it may also come about from the interplay between Hund's and Kondo interactions independent of the inter-site exchange (FM or AFM)[31].

Currently there is debate on whether the double phase transition in the specific heat at ambient pressure is really a manifestation of a two-component superconducting order parameter, as two thermodynamic transitions are not observed in all UTe$_2$ samples and single crystals with an optimal value of $T_c$ near 2 K exhibit only a single-phase transition[32,33]. The latter observation has led to the suggestion that the double transition at ambient pressure is due to an inhomogeneous distribution of two spatially separated regions of the sample with different $T_c$ values[34]. To complicate matters further, two well-defined transitions are induced by pressures above 0.3 GPa in single transition samples, with one phase transition increasing and the other decreasing with increasing pressure[35,36]. Even UTe$_2$ samples that display a single-phase transition are found to exhibit an anomalous extrapolated *T*-linear component in the specific heat $C(T)$ below $T_c$, as well as a ubiquitous upturn in $C/T$ vs $T$ at low temperatures[37]. The coefficient of the residual *T*-linear term ($\gamma^*$) varies among samples and is often a substantial fraction of the normal-state value ($\gamma_N$). Both the value of $T_c$ and the jump in the specific heat at the superconducting transition have been shown to increase with decreasing $\gamma^*/\gamma_N$[37,38].

While the low-temperature upturn in $C/T$ vs $T$ was suggested to arise from the high-temperature tail of a nuclear Schottky anomaly, the form of the upturn varies somewhat among samples[37,39]. The origins of the upturn and $\gamma^*$ in UTe$_2$ are unsettled issues.

Other heavy-fermion superconductors, including UPt$_3$, URu$_2$Si$_2$, UPd$_2$Al$_3$, and CeCoIn$_5$[40,41,42,43] also exhibit a large residual $T$-linear term in the low-temperature specific heat. Although some have proposed this to be an intrinsic property, $\gamma^*$ in these other materials is also sample dependent. One potential explanation for $\gamma^*$ is that there is a residual electronic contribution to $C(T)$ coming from some fraction of the sample that remains in the normal state. A residual $T$-linear term in the temperature dependence of the thermal conductivity $\kappa(T)$ would then be expected due to heat conduction by the unpaired 'normal' delocalized electrons. Another possibility is that $\gamma^*$ results from resonant impurity scattering in a superconducting state with a highly anisotropic order parameter[44]. A finite $T$-linear term in $\kappa(T)$ as $T \to 0$ is likewise expected, due to broadening of the gap nodes by impurity scattering resulting in a finite residual density of states at zero energy. Yet in contrast to other heavy-fermion superconductors, a vanishingly small residual $T$-linear contribution to the thermal conductivity of UTe$_2$ is observed[39]. This raises the possibility that the extrapolated $T$-linear component in $C(T)$ is of magnetic origin.

Here the magnetic properties of UTe$_2$ are addressed via a comprehensive µSR study of independently grown single crystals that exhibit a single or double phase transition in the specific heat. The investigation of multiple samples from independent sources enables us to distinguish common trends. We show that there is a fast-relaxing component of the µSR signal due to the occurrence of magnetic clusters and a slower relaxing component apparently due to spins that mediate interactions between the clusters. An additional large non-relaxing component in the µSR signal suggests there are negligible fields or fast spin fluctuations that persist down to at least ~ 0.02 K in a significant volume fraction of the sample. We find that percolation or growth of the magnetic clusters cease within the superconducting state, and the spin dynamics of the magnetic clusters slow down near the onset of superconductivity. Eventually all magnetic clusters freeze at lower temperatures where the upturn in $C/T$ vs $T$ occurs. We discuss how the magnetic clusters may account for the upturn and the residual $T$-linear term in the specific heat of UTe$_2$.

**Results**

Specific heat

Figure 1a-c shows the temperature dependence of the specific heat of a UTe$_2$ single crystal taken from each growth batch for the samples studied here by μSR, plotted as $C/T$ vs $T$. The specific heat data for sample S1 exhibits a double bulk phase transition and an upturn below ~ 0.3 K, whereas a single bulk phase transition is shown for samples S2 and S3 grown. Figure 1d shows $C/T$ vs $T$ data extending to lower temperatures for additional independently grown samples.

Zero field muon spin relaxation

The photos in Fig. 1 show the samples investigated by μSR mounted on silver (Ag) sample holders used for measurements utilizing a dilution refrigerator. Sample S1 is a randomly oriented mosaic of 22 irregularly shaped single crystals from the same growth batch. Samples S2 and S3 are flat, plate-like single crystals that were aligned with the *c*-axis perpendicular to the flat surfaces. Sample S3 is a lone single crystal, whereas S2 consists of a large single crystal from a different growth batch and two small single crystals from the same growth batch as S3.

Figure 2 shows zero-field (ZF) μSR spectra for samples S1 and S2 as a function of temperature. In contrast to the measurements on unaligned sample S1, the ZF spectra for samples S2 and S3 were recorded with the initial muon spin polarization **P**(0) parallel to the *c* axis. The ZF signal from muons stopping in the sample is well described by

$$a_0 P_z(t) = a_1 \exp(-\lambda_1 t) + a_2 \exp(-\lambda_2 t) + a_3, \tag{1}$$

where $P_z(t)$ is the time evolution of the muon spin polarization along the *z* axis, defined as the initial direction of **P**(0). The first two terms are temperature dependent and describe relaxation of the ZF signal by electronic moments in the sample. An exponential relaxation function $a_3 \exp(-\lambda_3 t)$ was initially used in place of the third term, but the fits for all three samples yielded a temperature-independent value of $\lambda_3 = 0.000 \pm 0.001$ μs$^{-1}$. There is also a non-relaxing temperature independent contribution to the measured ZF signal from muons stopping in the Ag sample holder that is not included in equation (1). The non-relaxing signals from muons stopping in the sample and the Ag sample holder cannot be distinguished in ZF. Hence the background contribution from muons stopping outside the sample was estimated from transverse-field (TF) measurements at $H = 20$ kOe (see Supplementary Note 1). The amplitudes and relaxation rates in the first two terms of equation (1) were found to play off each other in fits with all parameters free to vary with temperature. Consequently, $a_1$ and $a_2$ were considered as temperature-independent fit parameters for sample S1, while $a_1$ was left free to vary for samples S2 and S3 to account for an obvious reduction of the amplitude of the ZF signal with decreasing temperature, as explained below. These constraints in the data analysis are considered in the interpretation of the results.

The fits to equation (1) indicate a non-relaxing signal coming from a significant volume fraction of the sample, given by the ratio $a_3/a_0$ (Table 1). The typical relaxation of the ZF signal by randomly oriented nuclear moments is expected to be extremely weak in UTe$_2$. This is because the only stable uranium isotope with nonzero nuclear spin, depleted $^{235}$U, has a natural abundance of 0.20 % and the natural abundance of the tellurium isotopes with nuclear spin $^{123}$Te and $^{125}$Te are only 0.89 % and 7 %, respectively. Even so, a large temperature-independent non-relaxing signal from the sample is unexpected. It implies there are regions of the sample where either the root mean square of the local field at the muon site is extremely small or the local field fluctuates so fast that there is complete decoupling of the muon spin from the local field. In other words, even at the lowest temperature considered there are no electronic moments that are static or fluctuating slowly enough to cause a detectable relaxation of the ZF signal in a substantial portion of the sample volume.

We now consider the exponentially relaxing signals for all three samples. For the unoriented mosaic S1, approximately 27 % of the sample is described by a fast exponential relaxation rate ($\lambda_1$) and 37 % by a slower exponential relaxation rate ($\lambda_2$). Figure 2c shows the temperature dependence of $\lambda_1$ and $\lambda_2$ in sample S1, which exhibit behaviors resembling that observed in our earlier study of a different sample[21]. The temperature dependence of $\lambda_1/T$ overlaps with the earlier data above 0.75 K (Fig. 2c inset) where a fit of the current data to the relation $\lambda_1/T \propto T^{-n}$ yields $n = 1.43 \pm 0.06$ (see Supplementary Figure 3). While this is larger than the self-consistent renormalization (SCR) theory prediction for a 3-D metal near a FM instability[46] ($n = 1.33$), the value for S1 overlaps $n = 1.35 \pm 0.04$ determined in the earlier study. We note that the fast relaxation component is a smaller fraction of the total ZF signal compared to our previous study and consequently there is more scatter in the present data for $\lambda_1(T)$. Here it is now evident though that the change in the variation of $\lambda_1/T$ with $T$ at lower temperatures culminates with a saturation of $\lambda_2$ below ~ 0.2 K and a ZF signal with a single exponential relaxation component (*i.e.*, $\lambda_1 \sim \lambda_2$).

By contrast, there is a significant loss of the initial amplitude of the ZF signal for the *c*-axis aligned single crystals (samples S2 and S3) over the entire temperature range, with some remnant of a fast-relaxing component at early times that diminishes with decreasing temperature. The gradual loss of the fast-relaxing component in the initial dead time of the μSR spectrometer leads to a reduction of the amplitude $a_1$ and a large uncertainty in the fitted value of $\lambda_1$ (Fig. 2d). The more rapid relaxation rate in the *c*-axis aligned samples implies a wider distribution of local field sensed by the muon. This may result from a slower fluctuation rate and/or an anisotropic distribution of the local fields and suggests there are frozen internal magnetic fields already at 5 K.

The temperature dependence of $\lambda_2$ in samples S2 and S3 are quite similar. As in sample S1, $\lambda_2$ increases with decreasing temperature below 0.5 K and plateaus below ~ 0.12 K (Fig. 2d). There is also an abrupt change in the temperature dependence of $\lambda_2$ at $T_c$, with an initial saturation of $\lambda_2$ below $T_c$. The peak in $\lambda_2$ near $T_c$ for sample S3 indicates a significant change in the dynamics of the local magnetic fields. This peak is likely smeared out in sample S2, which consists of single crystals from two different growth batches exhibiting slightly different magnitudes and widths of the specific heat jump at $T_c$ (Fig. 1) and is entirely absent in sample S1.

Longitudinal field muon spin relaxation

Figure 3a-c shows the time evolution of the muon spin polarization in sample S1 recorded in a longitudinal-field (LF) experimental configuration with a magnetic field applied parallel to **P**(0) in three temperature regimes. At 9.5 K, there is very little variation in the relaxation of the LF signal between 100 Oe and 1 kOe, indicating the existence of rapidly fluctuating internal magnetic fields. At 0.3 K, the LF signals display a fast-relaxing front-end and a slow decaying "tail". The increasing amplitude with field of the long-time tail is indicative of slowly fluctuating local fields and the rapid relaxation at early times indicates a broad distribution of local fields.[47] As we explain later, these two

components come from different regions of the sample. At 0.06 K the long-time residual relaxation of the LF signal vanishes near 150 Oe, indicating the local fields associated with this component become quasistatic. Above 150 Oe the late-time amplitude of the LF signal moves up with increasing field due to decoupling of the rapid early-time relaxation. This is compatible with a broad distribution of quasistatic internal magnetic fields in a different portion of the sample.

Figure 3d shows the field dependence of the LF signal for sample S2 well above $T_c$. The applied LF increasingly brings signal amplitude back from the initial dead time as the LF approaches the characteristic width of the internal magnetic field distribution. At 5 K a LF of 10 kOe is insufficient to completely suppress the muon spin depolarization, suggesting there are rapidly fluctuating local fields in some fraction of the sample. At 0.06 K, there is no long-time relaxation of the LF signal at 500 Oe (Fig. 3e), which is indicative of quasistatic local fields. However, there is still some loss of signal amplitude with an early-time remnant of the fast-relaxing component. This observation is compatible with a wide distribution of quasistatic local fields. As shown in Fig. 3f, a rapidly damped small-amplitude remnant coherent precession signal is visible in the ZF signal of sample S2 at low temperatures, indicating the existence of magnetically ordered regions in the sample. The frequency of the oscillation corresponds to a local field of ~ 310 G. A similar oscillatory signal is not observed in samples S1 or S3 (see Supplementary Figure 2), which suggests that the local field distribution is sufficiently broad enough to prevent the development of a coherent precession signal or there are fewer magnetically-ordered regions in these samples.

Weak transverse field muon spin rotation

To gain further insight on the nature of the magnetic environment in UTe$_2$, measurements were performed in a weak transverse field (wTF), where the magnetic field is applied perpendicular to the initial muon spin polarization **P**(0). In a paramagnetic state the muon spin precesses around the wTF at the Larmor frequency $\omega = \gamma_\mu B_{\text{wTF}}$, where $\gamma_\mu$ is the muon gyromagnetic ratio and $B_{\text{wTF}}$ is the magnitude of the wTF. In the magnetically ordered environment of UTe$_2$ the local magnetic field greatly exceeds the wTF and the signal is rapidly depolarized. Figure 4a shows the time evolution of the muon spin polarization in sample S2 after cooling in a wTF of 23 Oe applied in the *a-b* plane perpendicular to **P**(0). There is a clear reduction in the amplitude of the wTF signal at the lower temperature, which occurs due to muons stopping in magnetic regions of the sample where they experience a wide distribution of local fields. The wTF signals were fit to the following two-component function

$$a_{\text{wTF}} P_z(t) = a_4 \exp(-\sigma_4^2 t^2)\cos(\omega_4 t + \phi) + a_5 \exp(-\sigma_5^2 t^2)\cos(\omega_5 t + \phi). \quad (2)$$

While fits to this function provide an accurate determination of the total amplitude $a_{\text{wTF}}$, the signals from the paramagnetic volume fraction of the sample and muons stopping outside the sample cannot be resolved at 23 Oe. Consequently, the ratio of the amplitude of the wTF signal from muons stopping outside the sample ($a_B$) and the maximum amplitude of the wTF signal is determined from the TF measurements at 20 kOe (see Supplementary Note 1). Since the full amplitude of the ZF signal in samples S2 and S3 is not completely recovered at 5 K, the maximum amplitude of the wTF signal was determined in separate calibration measurements on an Ag sample ($a_{\text{Ag}}$) mounted on the same sample holder used for each UTe$_2$ sample. The magnetic volume fraction of the sample is then calculated as $V_f = (a_{\text{Ag}} - a_{\text{wTF}})/(a_{\text{Ag}} - a_B)$. For consistency, the magnetic volume fraction of the sample associated with the missing amplitude in the ZF signal is calculated assuming the same values of $a_{\text{Ag}}$ and $a_B$. The calculated values of $V_f$ show good agreement between the wTF and ZF data (Fig. 4b). Although the wTF data for sample S3 is limited, it is clear this sample has a smaller magnetic volume than sample S2. We speculate that this is the reason it has a larger specific heat jump at $T_c$ (Fig. 1). The peak in the ZF data for $V_f$ near $T_c$ for sample S3 is highly correlated with the peak in $\lambda_2$ (Fig. 2d) and does not necessarily imply a peak in the magnetic volume fraction. Rather it is another indication of an abrupt change in the evolution of the magnetic regions in the sample. Similar wTF measurements were not carried out on sample S1, although the ZF data indicate a slight loss of the signal in the initial dead time (see Supplementary Figure 2), below ~ 0.3 K where $\lambda_1$ decreases and $C(T)$ increases with decreasing temperature. We note that our previous measurements[21] on an unoriented single-crystal mosaic of UTe$_2$ with a similar value of $T_c$ showed a larger reduction in the amplitude of the wTF signal at 0.25 K compared to the signal at 2.5 K.

## Discussion

Density functional theory (DFT) calculations suggest a single crystallographic muon site at (0.44, 0, 0.5) in $UTe_2$ (Fig. 5a, b). The DFT calculations do not consider local distortions of the crystal lattice by the positive muon, which may change this site slightly. For a single muon site in the absence of long-range magnetic order, the presence of relaxing and non-relaxing components in the ZF signal from the sample means that the magnetic properties of $UTe_2$ are spatially inhomogeneous. The non-relaxing component suggests there are regions of the sample where the U moments fluctuate so rapidly that they are completely decoupled from the muon spin, even at millikelvin temperatures. Alternatively, if there are quasistatic magnetic moments in these regions of the sample, the magnetic moments must be very small. From the uncertainty in the fitted value of the relaxation rate of the non-relaxing signal (0.001 $\mu s^{-1}$), we estimate the width of the distribution of any quasistatic local fields of nuclear or electronic origin in the non-relaxing regions of the sample to be less than 1 mT.

Previously, we attributed the presence of two exponential relaxing components in $UTe_2$ to two distinct crystallographic muon sites[21]. Since the DFT calculations suggest a single muon site and it is evident that the ratio of the two relaxation rates ($\lambda_1/\lambda_2$) is not independent of temperature, we consider other possibilities. In the temperature range (0.3-2 K) where two exponential relaxation rates are well resolved in sample S1, the presence of fast and slow relaxation rates is reminiscent of homogeneous spin freezing, albeit in a fraction of the sample volume. However, the amplitude of the slow-relaxing exponential "tail" of the ZF signal is greater than the amplitude of the fast-relaxing component ($a_2 > a_1$), which is not possible for randomly oriented local fields in a homogeneous spin freezing scenario and instead would imply a preferred orientation of the local field at the muon site in $(a_1 + a_2)/a_0$ of the sample volume (see Supplementary Note 3). Since no long-range magnetic order is observed in any of the samples, we instead attribute $\lambda_1$ and $\lambda_2$ to an inhomogeneous environment in which the fast-relaxing ZF signal is due to magnetic regions where frozen or very slow fluctuating spins generate a wide distribution of internal magnetic field and the slow-relaxing signal to paramagnetic regions of the sample where the electronic moments are fluctuating within the μSR time window above ~ 0.2 K.

The fast relaxation in all three samples is presumably associated with the development of extremely slow longitudinal magnetic fluctuations along the *a*-axis, detected as an initial rapid growth of the $^{125}$Te NMR spin-spin relaxation rate $1/T_2$ below 30-40 K for a magnetic field $H \parallel a$ axis[20]. Recently these slow magnetic fluctuations have been attributed to the growth of long-range FM correlations within the U-ladder sublattice structure[48]. In this same study, a breakdown in scaling between the normal-state bulk magnetic susceptibility and $^{125}$Te NMR Knight shift below ~ 10 K has been attributed to the formation of magnetic clusters by disorder/defect induced local disruptions of the long-range correlations, based on analogous behavior of low-dimensional correlated magnets with quenched disorder. Here we note that theoretically it has been shown that disorder in quasi 2-D and 3-D geometrically frustrated systems can also result in the formation of magnetic clusters of various sizes containing locally ordered spins[49,50]. In UTe$_2$ the combination of potential strong magnetic frustration imposed by the staggered two-leg ladder-type arrangement of the U atoms[29] (see Fig. 5c) and low-levels of disorder may induce the development of magnetic clusters.

A non-oscillating component in the wTF μSR signal is consistent with muons stopping in or near magnetic clusters where they experience strong internal magnetic fields along the muon spin polarization direction, much larger than the applied field. The observation of a spontaneous spin precession signal in sample S2 below ~ 0.2 K (Fig. 3f) provides further evidence of locally ordered spins and the more rapid depolarization of the ZF signal in the *c*-axis aligned single crystals is presumably caused by the anisotropic internal magnetic field distribution generated by the magnetic clusters. The non-relaxing component in the ZF signal indicates that correlations of electronic moments are either absent or undisturbed in a significant volume of all three samples. In the non-relaxing regions of the sample, FM or AFM fluctuations may be present that are simply too fast to be detectable in the μSR time window.

In contrast to NMR $1/T_2$ studies, our μSR measurements of the magnetic properties of UTe$_2$ extend well below $T_c$. The comprehensive wTF data for sample S2 (Fig. 4b) indicates a growth in the volume of magnetic clusters with decreasing temperature, which is halted within the superconducting state. The temperature dependence of $\lambda_2$ in samples S2 and S3 (Fig. 2d) appears correlated with the saturation of the magnetic volume fraction at $T_c$. While this behaviour is more subtle in sample S1, there is some initial saturation of $\lambda_1$ below $T_c$ (Fig. 2c) that suggests the fast and slow relaxing signals are not completely resolved near $T_c$ where $\lambda_2$ is quite small. Above ~ 0.5 K the temperature dependence of $\lambda_2$ in samples S2 and S3 closely resembles that of the relaxation rate of the wTF signal from muons stopping in paramagnetic regions of the FM cluster-phase of $(Fe_{1-x}Ni_x)_3GeTe_2$[51]. In the latter compound the relaxation rate of the paramagnetic regions increases as the temperature is lowered, peaks at a magnetic transition temperature where the growth in the volume of magnetic clusters ceases and saturates below the transition. Hence, we attribute $\lambda_2$ in UTe$_2$ to the width of the motionally-narrowed field distribution of independent spins outside the magnetic clusters that mediate interactions between them. The quantity $V_f$ is a measure of the volume fraction of the sample occupied by the magnetic clusters. The data presented in Fig. 4b shows that $V_f$ is smaller in sample S3, which exhibits the sharpest superconducting transition in the specific heat. However, we do not see evidence in the existing data that the saturation of $V_f$ or slowing down of the fluctuation rate of the magnetic clusters is responsible for the double transition observed in the specific heat of sample S1.

While our previous experiment[21] was consistent with the variation $\lambda_1/T \propto T^{-4/3}$ predicted for a 3-D metal in proximity to a FM instability, which is quite different from the prediction $\lambda_1/T \propto T^{-3/4}$ near an AFM instability[46], the critical exponent determined in the present study is somewhat higher than 4/3. As mentioned earlier, this may be a consequence of a reduced sensitivity to $\lambda_1$ due to a smaller fast-relaxing ZF component. However, since $\lambda_1$ is apparently associated with slow longitudinal magnetic fluctuations along the *a*-axis, the predictions for a 3-D metallic system are probably not relevant. Furthermore, the dynamics of magnetic clusters near a quantum phase transition can be quite different and in an itinerant spin system differ from power-law behavior[52]. Hence, we cannot tell from the current data whether the clusters are FM or AFM.

The low-temperature upturn in $C/T$ vs $T$ has been attributed to a nuclear Schottky anomaly associated with a crystal electric field (CEF) level splitting $\Delta E \sim 50$ mK of the $^{235}$U nuclear energy levels[37]. However, as argued previously[39] and shown here for sample S1, the upturn in some samples is far too broad for this explanation. In sample S1, the value $\Delta E \sim 317$ mK determined from a fit of the low-$T$ upturn to the analytical expression for a two-level Schottky anomaly (see Supplementary Note 4) is six times larger than the splitting expected due to the CEF. In some compounds the Schottky anomaly is attributable to Zeeman splitting of the degenerate ground state energy levels of the magnetic ion by the internal magnetic field generated by short-range magnetic correlations[53,54]. The non-universal Schottky anomaly observed in UTe$_2$ is then likely due to the presence of magnetic clusters that split the $^2F_{5/2}$ ground state of the U ions. In sample S2 where a local internal field of ~ 310 G occurs at the muon site due to magnetically ordered regions (Fig. 3f), a level splitting of $\Delta E \sim 50$ mK due to the net magnetic moment resulting from the magnetic clusters is $\mu^* = k_B \Delta E / 2B < 1.2 \, \mu_B$. While a larger value of $\mu^*$ may be required to explain the deduced value of $\Delta E \sim 317$ mK in sample S1, the analysis of the low-temperature upturn in $C/T$ vs $T$ does not account for broadening effects caused by the inhomogeneous nature of the cluster freezing.

Lastly, we consider the origin of the large residual $T$-linear term in the specific heat of UTe$_2$ below $T_c$. Initially this was considered intrinsic and due to a large residual density of states at the Fermi level, suggestive of a non-unitary spin-triplet pairing state[1,2]. Yet there is no residual fermionic heat transport observed in thermal conductivity measurements that could be attributed to a lingering density of states of itinerant quasiparticles in the superconducting state[39]. Furthermore, the value of $\gamma^*/\gamma_N$ varies significantly between samples. Here $\gamma^*/\gamma_N$ is significantly smaller in sample S3 compared to samples S1 and S2, and the ratio $(\gamma^*/\gamma_N)_{S3}/(\gamma^*/\gamma_N)_{S2}$ is similar to the ratio of the magnetic volume fractions $(V_f)_{S3}/(V_f)_{S2}$ in samples S2 and S3 (Table 1). Hence, the detection of disorder-induced slowly fluctuating magnetic clusters that ultimately freeze suggest that the residual $T$-linear term is at least in part magnetic in origin. A low-temperature $T$-linear term in $C(T)$ is a general property of spin glasses[55]. The broad internal field distribution sensed in our experiments is suggestive of clusters of locally ordered spins in UTe$_2$ behaving as spin-glass-like magnetic moments. When spin clusters are the basic entity, a $T$-linear term arises from inter-cluster interactions[56] and disordered spins between clusters[57]. The $T$-linear term observed in a spin-glass system normally occurs well below the freezing temperature. In UTe$_2$ the freezing of magnetic clusters is an inhomogeneous process that begins well above $T_c$ (Fig. 4b). Consequently, a $T$-linear term in $C(T)$ associated with spin freezing may occur well above the upturn in $C/T$ vs $T$.

The gradual freezing of magnetic clusters in the superconducting state has potential implications for other experimental findings in UTe$_2$ at low temperatures. These include (i) the finite magnetization detected via a magnetic-field-trained polar Kerr effect[18,22], (ii) an unexplained reduction in the intensity of the 1 meV resonance observed by INS below ~ 0.4 K[23], (iii) evidence of two-gap superconductivity from a low-temperature downturn in $1/T_1T$ measured by NMR[13] and (iv) additional peaks that appear in the $^{125}$Te NMR spectrum at low temperatures for $H \parallel b$ axis[58]. Our findings suggest that the intrinsic properties of UTe$_2$ will be more accessible in samples that display a negligible residual $T$-linear term and a modest Schottky anomaly in the specific heat below $T_c$.

**Methods**

Single crystal growth

The UTe$_2$ single crystals were grown at the University of Maryland (samples S1 and S5) and Los Alamos National Laboratory (samples S2, S3 and S4) by a chemical vapor transport method. Solid pieces of depleted uranium (99.99 %) and tellurium (Alfa Aesar, 99.9999 %) were weighed in a 2:3 ratio with total mass of ~ 1 g. The elements were sealed under vacuum using a hydrogen torch in a quartz tube along with ~ 0.02 g of iodine (J.T. Baker Inc., 99.99 %) in the case of sample S1 and ~ 0.2 g of iodine (Alfa Aesar, 99.99 %) for samples S2 and S3. A temperature gradient was maintained in a multi-zone furnace for 28 days (sample S1) and 11 days (samples S2, S3 and S4). The elements were placed in the hot end of the gradient at $T_i$, whereas single crystals of UTe$_2$ were obtained at $T_f$, the cold end of the gradient. For sample S1: $T_i$ = 1060 °C and $T_f$ = 1000 °C, sample S2: $T_i$ = 800 °C and $T_f$ =720 °C, sample S3: $T_i$ = 790 °C and $T_f$ =725 °C, and sample S4: $T_i$ = 800 °C and $T_f$ =725 °C. A modified method was used for the growth of sample S5 that will be reported elsewhere.

Specific heat measurements

The specific heat measurements of samples S1 and S4 were performed with an adiabatic heat pulse method. Due to a very short internal relaxation time constant the data for each temperature decay curve were fitted well with a single exponential expression. The specific heat measurements of samples S2 and S3 were performed to $^3$He temperatures using a Quantum Design calorimeter that utilizes a quasi-adiabatic thermal relaxation technique.

## Muon spin relaxation/rotation (μSR)

The μSR measurements were performed by implanting nearly 100 % spin-polarized positive muons (μ$^+$) into the sample and detecting the time evolution of the muon spin polarization $P_i(t)$ ($i = x, y, z$) via the muon decay positrons[47]. The zero-field (ZF) measurements were performed with the direction of the initial muon spin polarization **P**(0) anti-parallel to the muon beam momentum $p_u$ (defined to be in the $+z$-direction), and $P_z(t)$ measured by way of a pair of positron detectors arranged in a *forward-backward* ($\pm z$) geometry about the sample. The same experimental arrangement was used for the longitudinal-field (LF) measurements in which an external magnetic field was applied parallel to **P**(0) along the $z$-axis. The weak transverse-field (wTF) measurements of $P_z(t)$ were performed with an external magnetic field applied perpendicular (along the $x$-axis) to **P**(0). The high-TF measurements at $H$ = 20 kOe were performed with the external magnetic field applied parallel to the $z$-axis but with **P**(0) rotated to be along the $x$-axis perpendicular to the field. In the high-TF experiments, $P_x(t)$ was measured by way of a pair of positron detectors arranged in a left-right ($\pm x$) geometry about the sample.

The μSR measurements on all three UTe$_2$ samples were carried out using an Oxford Instruments top-loading dilution refrigerator on the M15 beamline at TRIUMF, Vancouver, Canada. The air-sensitive single crystals were first mounted on a pure Ag backing plate using a 50/50 mixture of Apiezon N and Cryocon grease in a glove box with an Argon overpressure. The backing plate was subsequently thermally anchored to the Ag sample holder of the dilution refrigerator. A 12.5 × 23 mm sample holder was used for the measurements on samples S1 and S2, while a smaller 5 × 17 mm sample holder was used for the S3 single crystal. The 22 irregular -shaped single crystals of sample S1 formed an unaligned mosaic, whereas the three single crystals of S2 and the lone single crystal of S3 were mounted with the *c*-axis aligned parallel to the initial muon spin polarization **P**(0) in the ZF, LF and wTF experiments, and perpendicular to **P**(0) (but parallel to the applied field) in the high-TF experiments.

## Density functional theory (DFT) calculations

To find the muon site in UTe$_2$, DFT calculations were performed with the QUANTUM ESPRESSO package[59]. These calculations were performed using a 9 × 6 × 3 Monkhorst-Pack grid, and ultrasoft pseudopotentials were used to model the electronic wavefunctions of the atoms. Specifically, the 5*s* and 5*p* electrons were treated as valence electrons for Te, and the 6*s*, 7*s*, 6*p*, 6*d* and 5*f* electrons were treated as valence for U. The strong spin-orbit coupling in U was also included in the calculation, as was 0.01 Ry of Gaussian smearing. The kinetic energy cutoff used was 85 Ry, and the charge density cutoff was set to 850 Ry.

The total potential, consisting of the sum of the bare ionic, the exchange and correlation, and the Hartree potentials, is shown in Fig. 5a, b. The sign of the potential is chosen so that the positively charged muon is most likely to go to the position of the maximum of this potential, and as such these results show that the most likely muon stopping site is (0.44, 0, 0.5). Muons are known to distort their local environment, which may result in the actual muon site being slightly displaced from this[60,61,62]. Evaluating the extent of these distortions via a DFT relaxation calculation has proven difficult due to the larger number of electrons of the U atom and is an ongoing effort.

**Acknowledgements**

J.E.S. and S.R.D. acknowledge support from the Natural Sciences and Engineering Research Council of Canada. Research at the University of Maryland was supported by the Department of Energy Award No. DE-SC-0019154 (low temperature experiments), the Gordon and Betty Moore Foundation's EPiQS Initiative through Grant No. GBMF9071 (materials synthesis), NIST, and the Maryland Quantum Materials Center. S.R.S. acknowledges support from the National Institute of Standards and Technology Cooperative Agreement 70NANB17H301. S.J.B. and J.M.W. acknowledge support from EPSRC (Grant No. EP/ N023803/1). The DFT+$\mu$ calculations were performed on the Redwood Cluster at the University of Oxford. Work at Los Alamos National Laboratory (LANL) was performed under the auspices of the U.S. Department of Energy, Office of Basic Energy Sciences, Division of Materials Science and Engineering project "Quantum Fluctuations in Narrow-Band Systems." S.L. and A.W. acknowledge support from the Laboratory Directed Research and Development program at LANL.

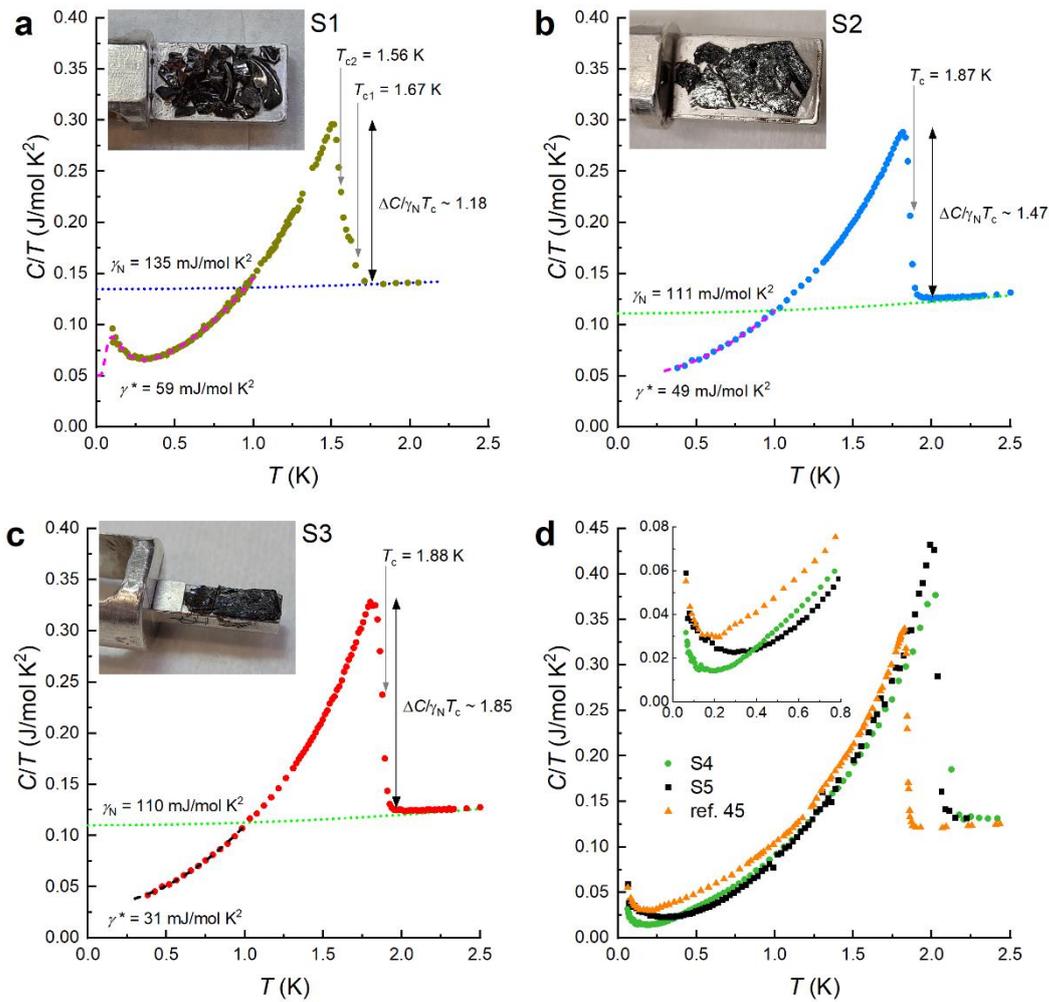

**Fig. 1 | Zero-field specific heat of the UTe$_2$ single crystals as a function of temperature.** Samples **a**, S1 **b**, S2, and **c**, S3 studied by μSR. The dotted curves are fits of the specific heat ($C$) data (plotted as $C/T$ vs $T$) above the superconducting transition temperature $T_c$ to $C/T = \gamma_N + \alpha T^2$, where $\gamma_N$ is a constant residual term and $\alpha$ is a coefficient of the $T^2$ term. The dotted curves for samples S2 and S3 come from fits of data up to 5 K (Not shown). The dashed curves are fits below 1 K to $C/T = \gamma^* + \beta T^2 + \alpha(\Delta E/k_B)^2 \exp(\Delta E/k_B T)[1 + \exp(\Delta E/k_B T)]^{-2} T^{-3}$, where $\gamma^*$ is a constant residual term and $\beta$ is a coefficient of the $T^2$ term. The last term is the analytical relation for the Schottky anomaly of a two-level system with an energy level splitting $\Delta E$, where $\alpha$ is a temperature-independent prefactor and $k_B$ is Boltzmann's constant. This term vanishes in the fits of the existing data for samples S2 and S3, whereas the fit for sample S1 yields $\Delta E = 317 \pm 9$ mK (see Supplementary Note 4). The photos in **a**, **b**, and **c** show the samples attached to the Ag sample holders used for the μSR measurements. The photo in **b** shows two pieces of a large S2 single crystal that broke and two small single crystals in the upper left from the same growth batch as S3. The photo in **c** is a lone S3 single crystal mounted on a smaller Ag sample holder. This crystal also broke when mounted. The fitted values of $\gamma^*$ for samples S1, S2 and S3 are shown in Table 1. **d**, Additional specific heat data for independently grown samples.

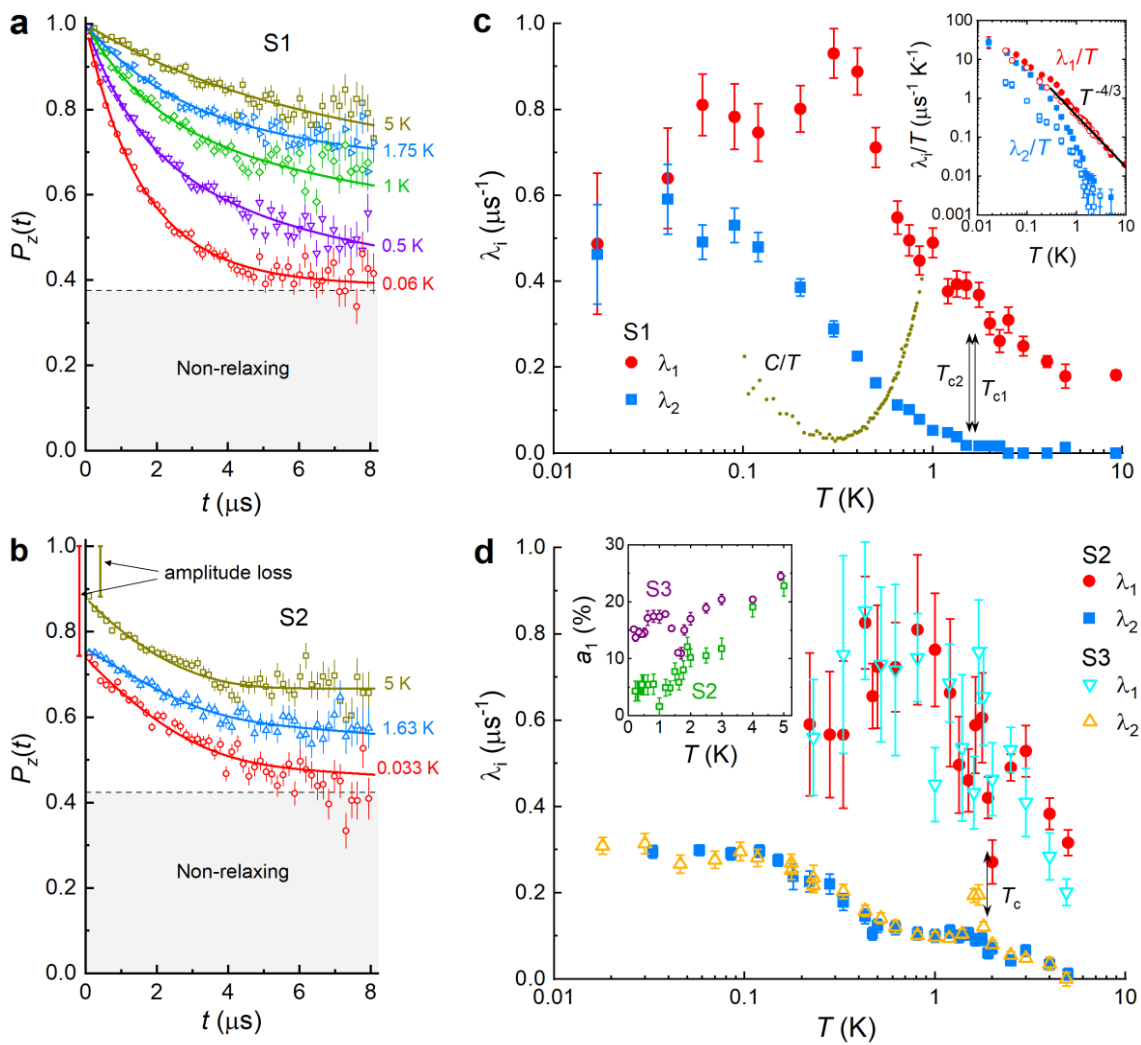

**Fig. 2 | Muon spin relaxation rate of UTe$_2$ single crystals in zero external magnetic field. a**, Representative zero-field (ZF) muon spin relaxation spectra of samples S1 and **b,** S2 for various temperatures after subtracting the non-relaxing background contribution from the signals. The solid curves in **a** and **b** are fits to equation (1). The size of the non-relaxing contribution to the spectra from the sample is indicated by the shaded regions. For sample S2 there is a loss of initial amplitude as indicated. **c,** Temperature dependence of the ZF exponential relaxation rates ($\lambda_1$ and $\lambda_2$) for S1 from fits of the ZF spectra to equation (1). Also shown is the $C/T$ vs $T$ data for sample S1 below 0.9 K, highlighting the low-temperature upturn (For visual clarity the vertical scale of the specific heat data is not shown). The inset shows that the temperature dependence of $\lambda_1/T$ and $\lambda_2/T$ for sample S1 (solid symbols) compared to the sample investigated in our earlier study[21] (open symbols). The solid line shows the proportional relationship $\lambda_1/T \propto T^{-4/3}$ predicted by the self-consistent renormalization theory for a 3-D metal near a ferromagnetic instability[46]. **d,** Temperature dependence of the ZF exponential relaxation rates for samples S2 and S3. The inset shows the temperature dependence of the remnant amplitude $a_1$ of the fast-relaxing component (as a percentage of the amplitude of the sample contribution to the ZF signal). Below 0.2 K a large fraction of the fast-relaxing signal is lost in the initial dead time. The error bars represent the standard deviation of the fit parameters.

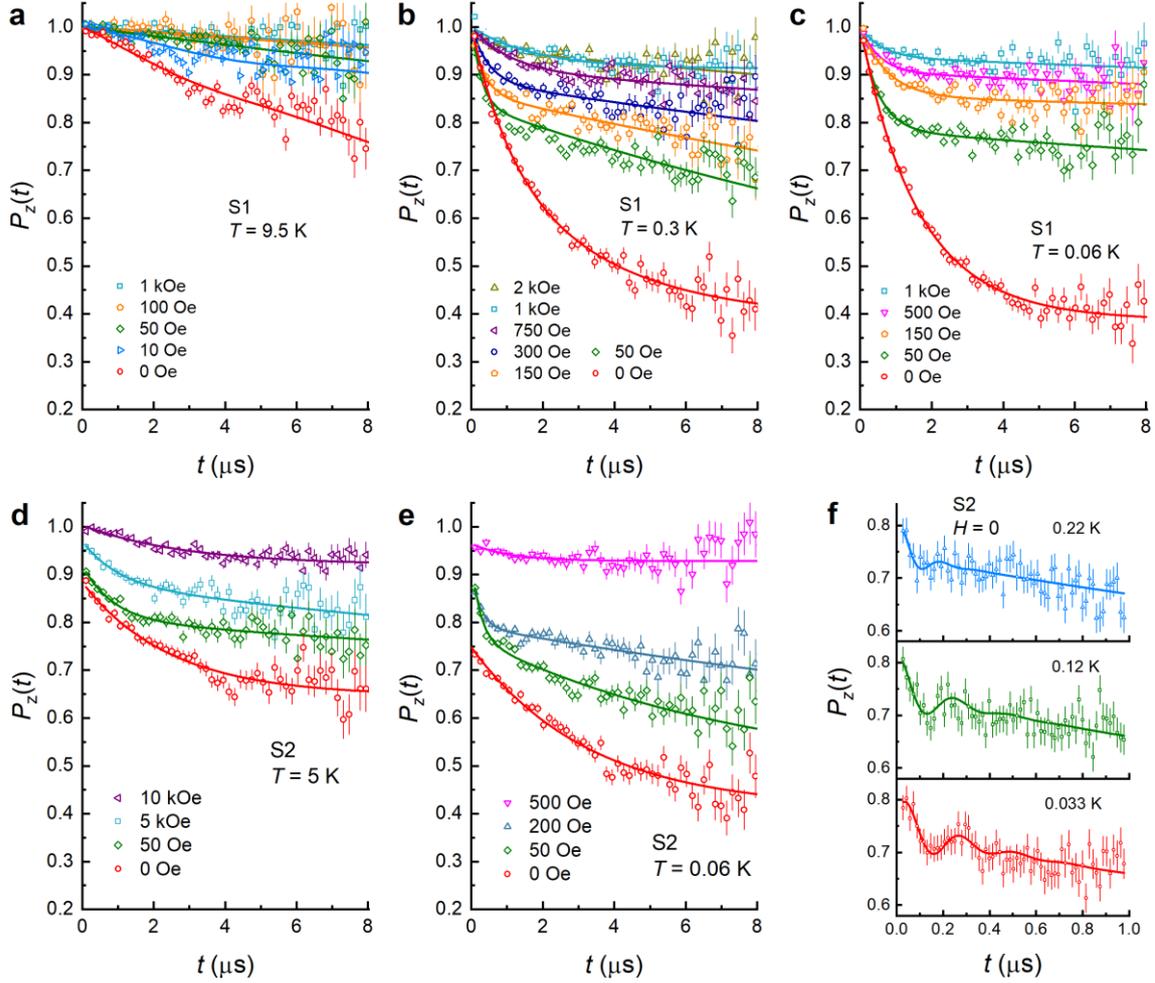

**Fig. 3 | Evidence for a nearly spin-frozen state.** Longitudinal-field (LF) muon spin relaxation spectra after subtracting the non-relaxing background contribution, recorded for sample S1 at **a,** 9.5 K, and below the superconducting transition temperature $T_c$ at **b,** 0.3 K and **c,** 0.06 K for various applied magnetic fields. Similar LF spectra for sample S2 at **d,** 5 K and below $T_c$ at **e,** 0.06 K for various magnetic fields applied parallel to the $c$-axis. The solid curves in **b**-**e** are fits to equation (1) and the curves in **a** are fits to equation (1) but with the exponential relaxation function in the second term replaced by a LF Gaussian Kubo-Toyabe function[47] to account for the relaxation by nuclear dipole fields. **f,** Zero-field (ZF) signals in sample S2 at early times with the non-relaxing background contribution removed. The solid curves are fits to equation (1) but with the first term multiplied by $\cos(\gamma_\mu B t + \phi)$, where $\gamma_\mu$ is the muon gyromagnetic ratio, $\phi$ is the phase shift and $B$ is the magnitude of the average local magnetic field at the muon site. The oscillatory component is indicative of short-range magnetic order.

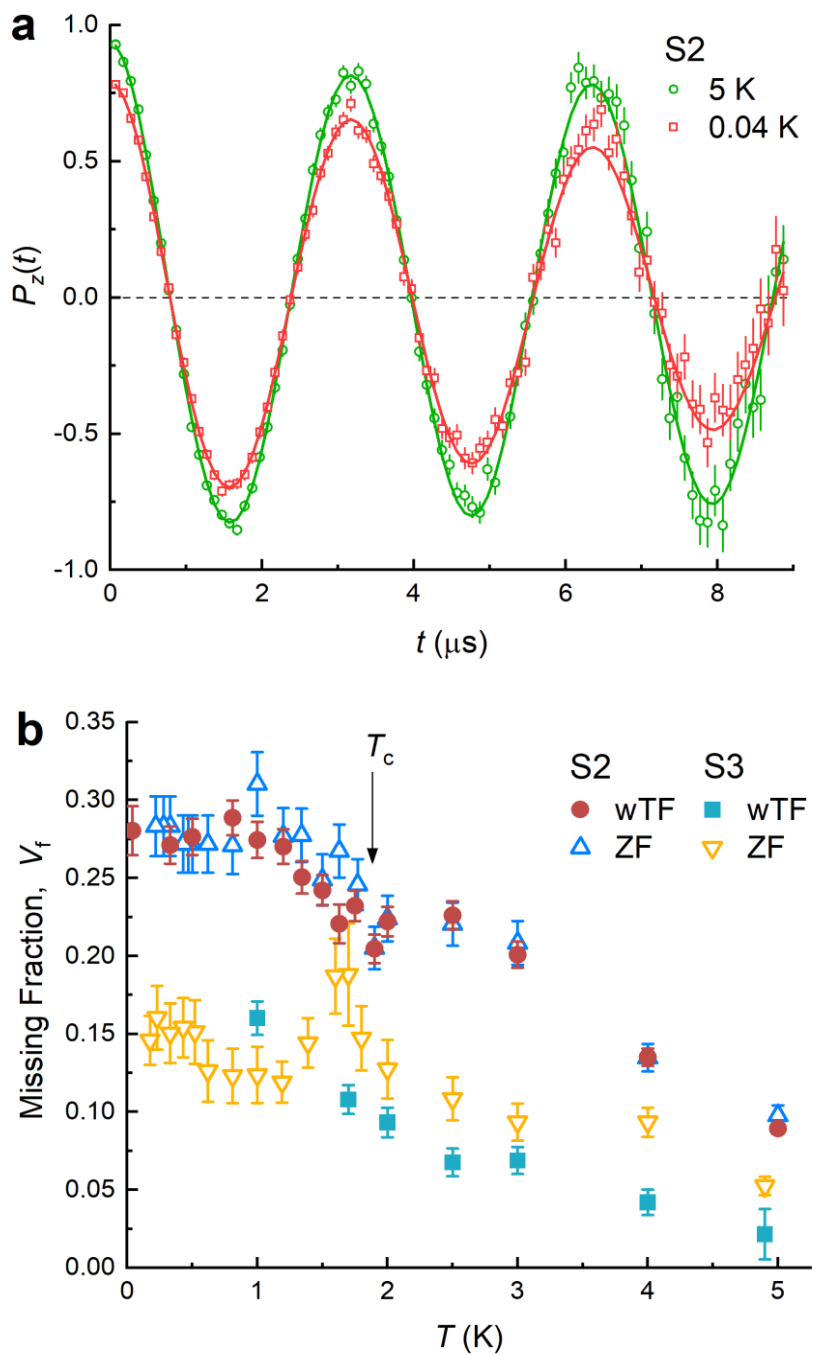

**Fig. 4 | Magnetic volume fraction in *c*-axis aligned UTe₂ single crystals exhibiting a single-phase transition in the specific heat. a**, Time evolution of the muon spin polarization for sample S2 recorded at 5 K and 0.04 K for a weak magnetic field $B_{\text{ext}} =$ 23 Oe applied in the *a-b* plane perpendicular to the initial muon spin polarization **P**(0). A portion of the signal is completely depolarized in the initial dead time of the spectrometer. The solid curves represent fits to equation (2). **b,** Temperature dependence of the volume fraction of samples S2 and S3 responsible for the loss of asymmetry in the weak transverse-field (TF) and zero-field (ZF) measurements. The error bars represent the standard deviation of the fit parameters.

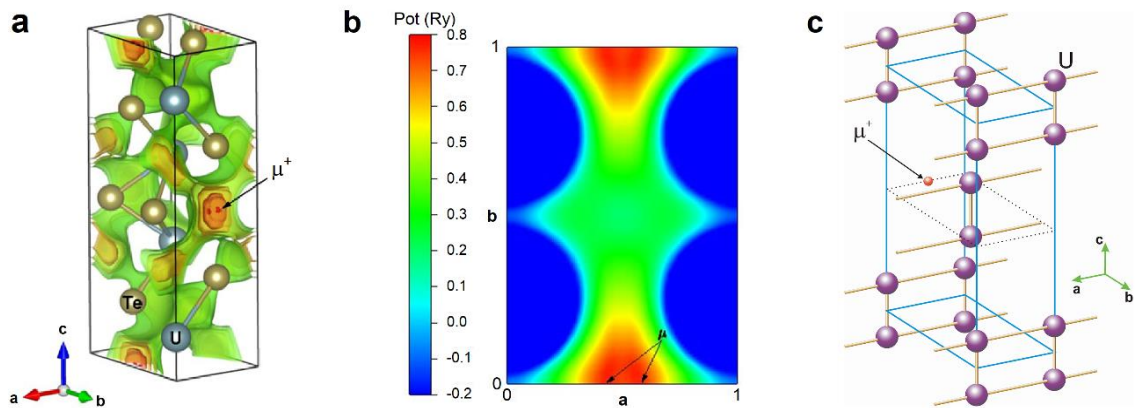

**Fig. 5 | Muon site in UTe$_2$. a**, Contours of the total potential energy calculated by density functional theory, and **b**, this same potential in the (0, 0, 2) plane. The energies in both figures are shaded according to the legend in **b**, although energies lower than 0.4 Ry (where Ry is the Rydberg unit of energy) are unshaded in **a** for visual clarity. The calculated muon site at (0.44, 0, 0.5) is indicated and in **b** this is indicated along with the crystallographic equivalent site at (0.56, 0, 0.5). **c**, The staggered two-leg ladder-type arrangement of the U atoms in UTe$_2$. Note the 'ladder legs' are parallel to the *a*-axis. For visual clarity the Te atoms are not shown. The small orange sphere indicates the calculated muon site at (0.44, 0, 0.5).

**Table 1 Comparison of specific heat and μSR parameters.**

|  | S1 | S2 | S3 |
|---|---|---|---|
| $T_c$ (K) | — | 1.87 | 1.88 |
| $T_{c1}, T_{c2}$ (K) | 1.67, 1.56 | — | — |
| $\Delta T_c$ (K) | 0.04, 0.05 | 0.05 | 0.05 |
| $\gamma_N$ (mJ/mol K$^2$) | 135 | 111 | 110 |
| $\gamma^*$ (mJ/mol K$^2$) | 59 | 49 | 31 |
| $\gamma^*/\gamma_N$ | 0.44 | 0.44 | 0.28 |
| $\Delta C/\gamma_N T_c$ | 1.18 | 1.47 | 1.85 |
| $a_1/a_0$ | 0.27 | — | — |
| $a_2/a_0$ | 0.37 | 0.23 | 0.24 |
| $a_3/a_0$ | 0.36 | 0.43 | 0.46 |
| $V_f$ | — | 0.28 | 0.16 |

The superconducting transition temperature $T_c$ of samples S2 and S3, the double phase transition temperatures $T_{c1}$ and $T_{c2}$ of sample S1, the phase transition widths $\Delta T_c$, the coefficient of the $T$-linear term in $C(T)$ above and below $T_c$ ($\gamma_N$ and $\gamma^*$), the specific heat jump at the superconducting transition $\Delta C/\gamma_N T_c$, the ratio of the amplitudes of the three components in Eq. (1), $a_1$, $a_2$, and $a_3$, to the total sample amplitude $a_0$, and the magnetic volume fraction $V_f$ at low temperatures.

# Supplementary Information

**Supplementary Note 1:** High transverse-field (TF) measurements

Transverse-field (TF) μSR asymmetry spectra for the three UTe$_2$ samples in an applied magnetic field of $H$ =20 kOe and at a temperature above $T_c$ are presented in Supplementary Figure 1a.

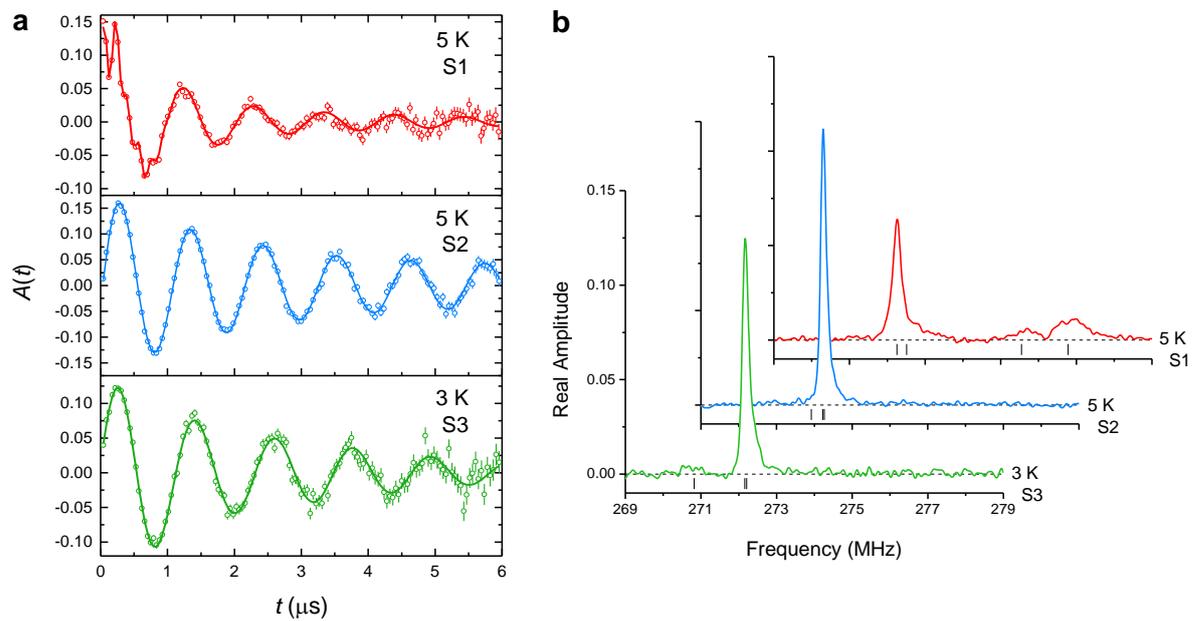

**Supplementary Figure 1. a,** TF-μSR asymmetry spectra for an applied magnetic field of $H$ = 20 kOe recorded at temperatures above $T_c$ displayed in a rotating reference frame (RRF) frequency of 271.3 MHz. The solid curves represent fits of the asymmetry spectra to Supplementary Equation (1) displayed in the same RRF. **b,** Fourier transforms (with Gaussian apodization) of the TF-μSR asymmetry spectra, which provide visual depictions of the internal magnetic field distribution $n(B)$ sensed by the muon ensemble. The frequency (horizontal axis) is related to the local internal magnetic field via the relation $f = (\gamma_\mu/2\pi)B$, where $\gamma_\mu = 8.5162 \times 10^8$ s$^{-1}$ T$^{-1}$ is the gyromagnetic ratio of the muon. The vertical black lines denote the average frequencies of the multiple components that describe the TF-μSR spectra.

The solid curves through the data points in Supplementary Figure 1a are fits to

$$A(t) = a_B \exp(-\Delta_B^2 t^2) \cos(2\pi f_B t + \phi) + \sum_{i=1}^{n} a_i \exp(-\Delta_i^2 t^2) \cos(2\pi f_i t + \varphi). \quad (1)$$

The first term is a background component due to muons that missed the sample and stopped in the Ag sample holder. Here $a_B$ is the amplitude of the background signal, $\phi$ is the angle between the axis of the positron detector and the initial muon-spin polarization **P**(0), and $f_B$ is the precession frequency of the muon spin in the magnetic field **B**. The Gaussian relaxation function $\exp(-\Delta_B^2 t^2)$ is intended to account for the distribution of internal magnetic field, associated with randomly oriented nuclear moments in the sample holder, although the value of $\Delta_B$ is influenced somewhat by dipolar fields emanating from the sample associated with the field-induced polarization of the local U electronic moments. The second term, which is a sum of *n* similar Gaussian damped cosine functions, describes the TF-μSR signal from muons stopping in the sample. The sample component of the signal for UTe$_2$ samples S1, S2 and S3 is comprised of *n* = 3, 2, and 2 components, respectively. The fitted parameters are listed in Supplementary Table 1.

| Sample | Frequencies (MHz) | Amplitudes (%) | Relaxation Rates (μs$^{-1}$) |
|---|---|---|---|
| S1 | $f_B$ = 272.255(2) | $a_B$ = 34(1) | $\Delta_B$ = 0.44(1) |
|    | $f_1$ = 272.51(3)  | $a_1$ = 31(1) | $\Delta_1$ = 2.2(1) |
|    | $f_2$ = 275.55(4)  | $a_2$ = 13.0(1) | $\Delta_2$ = 1.8(2) |
|    | $f_3$ = 276.78(2)  | $a_3$ = 22(1) | $\Delta_3$ = 1.57(8) |
| S2 | $f_B$ = 272.221(2) | $a_B$ = 30(1) | $\Delta_B$ = 0.05(1) |
|    | $f_1$ = 271.92(5)  | $a_1$ = 21(1) | $\Delta_1$ = 3.6(2) |
|    | $f_2$ = 272.261(5) | $a_2$ = 49(1) | $\Delta_2$ = 0.50(4) |
| S3 | $f_B$ = 272.1558(2) | $a_B$ = 40(2) | $\Delta_B$ = 0.20(1) |
|    | $f_1$ = 270.82(6)  | $a_1$ = 13(2) | $\Delta_1$ = 2.0(3) |
|    | $f_2$ = 272.21(1)  | $a_2$ = 47(2) | $\Delta_2$ = 0.80(3) |

**Supplementary Table 1**. Parameters from fits of the TF-μSR asymmetry spectra in Supplementary Figure 1a to Supplementary Equation (1).

Fourier transforms of the TF-μSR signals, which provide an approximate visual representation of the internal magnetic field distribution sensed by the muon ensemble, are shown in Supplementary Figure 1b. The additional broad higher-frequency peaks for sample S1 are presumably due to the misalignment of the single crystals in the mosaic, which results in different orientations of the magnetic dipole field from the local U moments in the individual crystals.

## Supplementary Note 2: Zero-field (ZF) measurements

### ZF signals at early times

Supplementary Figure 2 shows a comparison of the sample contribution to the low-temperature ZF-μSR spectra of samples S1, S2 and S3 at early times. All three samples display some loss of amplitude in the dead time of the spectrometer, such that $P_z(0) < 1$. A remnant oscillating component is evident only in sample S2.

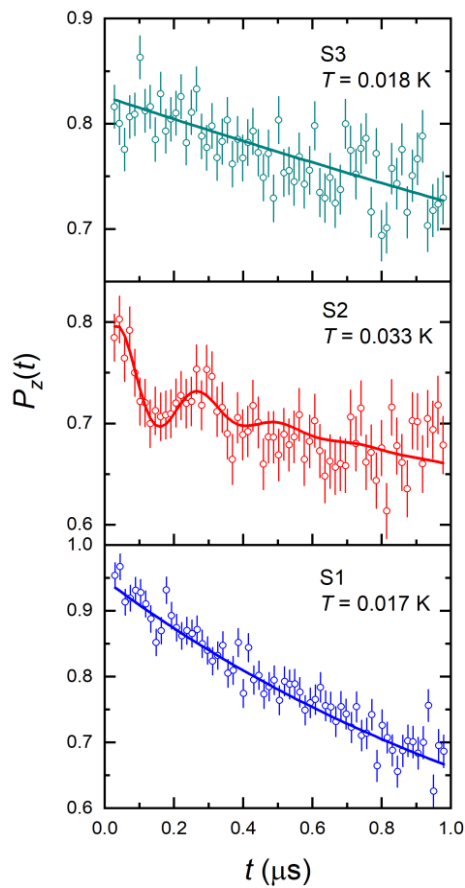

**Supplementary Figure 2.** ZF-μSR spectra of samples S3, S2 and S1 at early times and low temperatures after subtracting the non-relaxing background contribution from the signals. The solid curves are fits to equation (1) in the main text, but with the first term multiplied by $\cos(\gamma_\mu B t + \phi)$, where $\gamma_\mu$ is the muon gyromagnetic ratio, $\phi$ is the phase shift and $B$ is the magnitude of the average local magnetic field at the muon site. The fits yield $B = 311.8 \pm 74$ G for sample S2, and $B = 0$ G for samples S1 and S3.

### Fits of $\lambda_1/T$ versus $T$ for sample S1

Supplementary Figure 3a shows the fit of $\lambda_1/T$ versus $T$ above 0.4 K to the relation $\lambda_1/T \propto T^{-n}$ from our previous ZF-μSR study of UTe$_2$. Supplementary Figure 3b shows a comparison of $\lambda_1/T$ versus $T$ between the current results for sample S1 and the previous study of a different sample. There is good agreement between the two data sets above 0.75 K. The divergence of the two data sets below 0.75 K signifies a difference in the range of temperature over which the spins (magnetic clusters) freeze. Supplementary Figure 3c shows a fit of $\lambda_1/T$ versus $T$ for sample S1 above 0.75 K to the relation $\lambda_1/T \propto T^{-n}$.

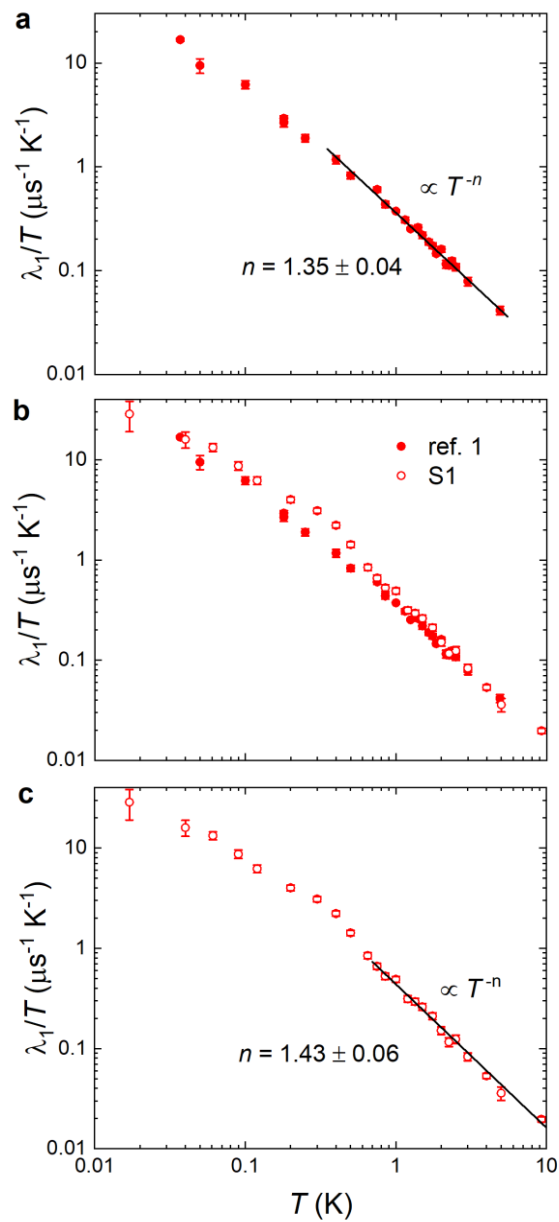

**Supplementary Figure 3. a,** Fit of $\lambda_1/T$ versus $T$ above 0.4 K to the relation $\lambda_1/T \propto T^{-n}$ from Supplementary Reference 1. **b,** Comparison of $\lambda_1/T$ versus $T$ for sample S1 to data from Supplementary Reference 1. **c,** Fit of $\lambda_1/T$ versus $T$ for sample S1 above 0.75 K to the relation $\lambda_1/T \propto T^{-n}$.

## Supplementary Note 3: Interpretation of $a_2 > a_1$ in the case of homogeneous spin freezing

The amplitude of the slow-relaxing component in the ZF signal ($a_2$) is larger than the amplitude of the fast-relaxing component ($a_1$). In the case of homogeneous spin freezing and a single muon stopping site, the time evolution of the muon spin polarization in a single crystal at zero field is given by[46]

$$P_z(t) = G_x(t)\sin^2\theta\cos(\gamma_\mu Bt) + G_z(t)\cos^2\theta, \tag{2}$$

where $\theta$ is the angle between the initial muon spin direction, $G_z(t)$ is a longitudinal relaxation function associated with the local field component parallel to the *z*-direction and $G_x(t)$ is a transverse relaxation function associated with the local field components perpendicular to the *z*-direction. While $G_z(t)$ is sensitive only to dynamic depolarization, $G_x(t)$ is sensitive to both static and dynamic effects. Consequently, $G_x(t)$ usually relaxes much faster with time than $G_z(t)$ and the static depolarization can be so strong that all or some of the $G_x(t)$ component can be lost in the initial deadtime of the spectrometer. For these reasons, if the relaxing part of the ZF signal in UTe$_2$ is due to homogeneous spin freezing, $\lambda_1$ (and hence $a_1$) is associated with $G_x(t)$ and $\lambda_2$ (and hence $a_2$) is associated with the more slowly decaying function $G_z(t)$. For randomly oriented local fields, $a_1 = 2/3$ and $a_2 = 1/3$, which is contrary to the experimental result $a_2 > a_1$. If on the other hand there is a preferred orientation of the local field (*i.e.*, such that $\theta$ is the same throughout the sample), then $a_2 > a_1$ if $\cos^2\theta > \sin^2\theta$.

## Supplementary Note 4: Specific Heat Measurements

**Fits of the low-temperature specific heat data of samples S1 and S4**

Supplementary Figure 4 shows fits of $C/T$ versus $T$ of samples S1 and S4 below 1 K, and Supplementary Tables 2 and 3 list the fit results. The fit to the high-temperature tail of a Schottky anomaly for a two-level system does not adequately describe the data for sample S1, whereas the fit assuming the full analytical expression for the Schottky anomaly adequately describes the specific heat data of both samples.

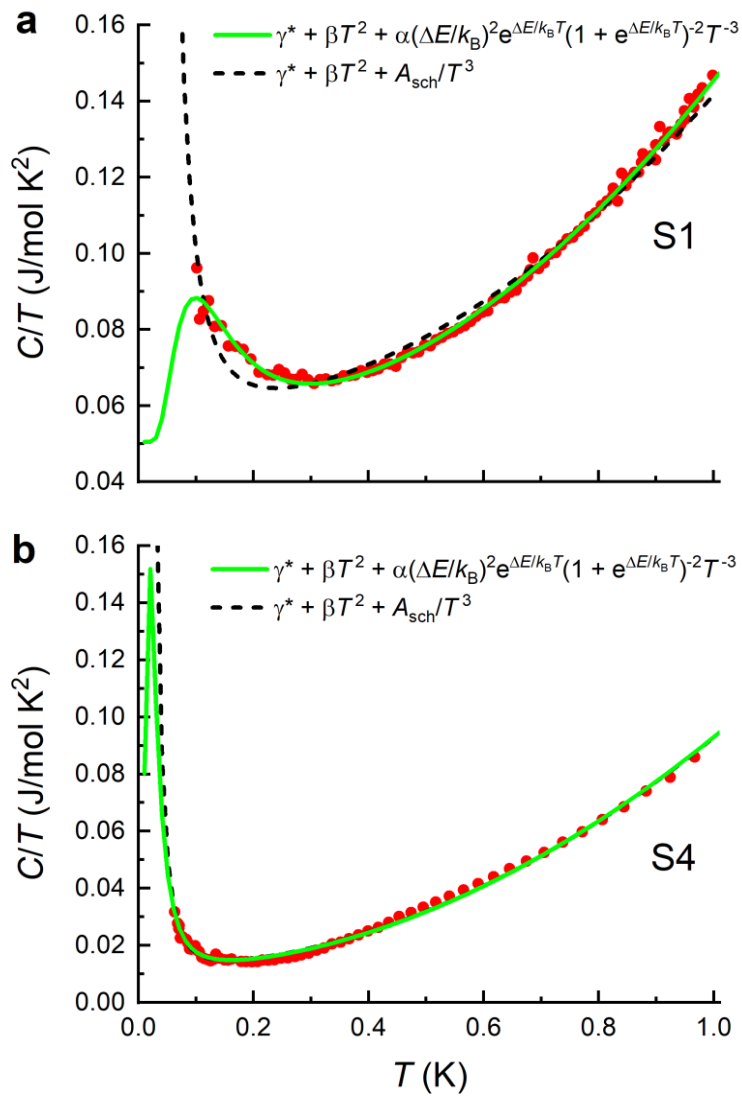

**Supplementary Figure 4.** Fits of $C/T$ versus $T$ for samples **a,** S1 and **b,** S4 assuming the high-temperature tail $A_{\text{sch}}/T^3$ (black dashed curves) and the full analytical relation for the Schottky anomaly of a two-level system (green solid curves).

| Sample | $\gamma^*$ (mJ/mol$^{-1}$ K$^2$) | $\beta$ (mJ/mol$^{-1}$ K$^{-4}$) | $A_{\text{sch}}$ (µJ/mol$^{-1}$ K) |
|---|---|---|---|
| S1 | $56.6 \pm 0.7$ | $85 \pm 2$ | $44 \pm 2$ |
| S4 | $11.6 \pm 0.2$ | $81.2 \pm 0.7$ | $5.0 \pm 0.2$ |

**Supplementary Table 2**. Results of fits of the specific heat data in Supplementary Figure 4 to the high-temperature tail of the Schottky anomaly for a two-level system: $C = \gamma^* T + \beta T^3 + A_{\text{sch}}/T^2$, where $\gamma^*$, $\beta$ and $A_{\text{sch}}$ are temperature-independent coefficients.

| Sample | $\gamma^*$ (mJ/mol$^{-1}$ K$^2$) | $\beta$ (mJ/mol$^{-1}$ K$^{-4}$) | $\Delta E/k_B$ (mK) |
|---|---|---|---|
| S1 | $50.5 \pm 0.5$ | $95 \pm 1$ | $317 \pm 9$ |
| S4 | $11.4 \pm 0.3$ | $81.6 \pm 0.8$ | $59 \pm 24$ |

**Supplementary Table 3**. Results of fits of the specific heat data in Supplementary Figure 4 to the full analytical relation for the Schottky anomaly of a two-level system: $C = \gamma^* T + \beta T^3 + \alpha(\Delta E/k_B T)^2 \exp(\Delta E/k_B T)[1 + \exp(\Delta E/k_B T)]^{-2}$, where $\gamma^*$, $\beta$ and $\alpha$ are temperature-independent coefficients, $\Delta E$ is the energy gap between the two levels and $k_B$ is Boltzmann's constant.

## Supplementary References